\documentclass[preprint,aps,showpacs,floatfix]{revtex4}
\usepackage{amsmath,amssymb,graphicx,epsfig}

\newcommand{\Rdot}{\dot{R}}
\newcommand{\zdot}{\dot{z}}
\newcommand{\udot}{\dot{u}}
\newcommand{\vdot}{\dot{v}}

\newcommand{\bkappa}{{\mbox{\boldmath $\kappa$}}}
\newcommand{\ukappa}{{\mbox{\boldmath $\hat \kappa$}}}
\newcommand{\rL}{\mathbf {r}_L}
\newcommand{\vL}{\mathbf {v}_L}
\newcommand{\vs}{\mathbf {v}_s}
\newcommand{\vi}{\mathbf {v}_i}
\newcommand{\vst}{\mathbf {v}_{st}}
\newcommand{\vn}{\mathbf {v}_n}

\begin{document}

\title{\bf Motion of vortex ring with tracer particles in
superfluid helium}

\author{Carlo F. Barenghi$^1$ and Yuri A. Sergeev$^2$}

\affiliation {
$^1$School of Mathematics and Statistics,
Newcastle University, Newcastle upon Tyne, NE1 7RU, England, UK\\
$^2$School of Mechanical and Systems Engineering,
Newcastle University, Newcastle upon Tyne, NE1 7RU, England, UK}

\date {\today}

\begin {abstract}
Recent experiments on quantum turbulence in superfluid helium
made use of small tracer particles to track the motion of quantized
vortices, determine velocity
statistics and visualize vortex reconnections. A problem with this 
visualization technique is that it may change the turbulent flow
which it visualizes. To address this problem,
we derive and solve the equations of motion of a quantized vortex 
ring which contains a given number of particles trapped on 
the vortex core, hence
derive a quantitative criterion to
determine in which measure small particles trapped in quantized
vortices affect vortex motion. Finally we apply the criterion to
a recent experiment.
\end{abstract}



\maketitle

\section{Motivation} 

The study of quantized vortices \cite{Donnelly} in superfluid helium 
(He~II) and of
quantum turbulence \cite{Walmsley2008,Roche2007,Bradley2008,
Chagovets2007,Eltsov2007,Vinen2002,B2008} has been held back over the years by 
the difficulty of flow visualization near absolute zero.
Fortunately the problem has been recognized, and new
visualization techniques have become recently available. One of the most
promising technique is based on trapping small micron-size tracer particles
(made of glass, polymers or solid hydrogen)
onto quantized vortices~\cite{PIV,Paoletti-JPSJ}. The use of tracer
particles has made possible, for example,
to study velocity statistics~\cite{Paoletti-stats} in superfluid
turbulence and to visualize~\cite{Bewley-reconnections} 
individual reconnections of quantized vortices, a process which is
crucial in the dynamics of turbulence~\cite{Alamri2008}. 

To interpret these recent experiments it is necessary to understand the
interaction of the tracer particles with the quantized vortices~\cite{Poole,Sergeev,Kivotides}. 
In particular, we need to find in which
measure the tracer particles disturb the vortices, hence change the flow
which one tries to visualize. Clearly, on one hand a large number
of tracers improves the images and the signal to analyze, but
on the other hand too many tracers must affect the motion of vortices
in a significant way.  What is the maximum density of tracers
which can be used?

To answer this practical question, we 
introduce a quantitive measure of the tracers' disturbance,
which we call the superfluid Stokes number. 
After deriving and solving the governing equations,
we determine by what amount the motion of
a quantized vortex ring of radius $R$ is disturbed by the presence
of $N$ tracer particles of radius $a$ trapped into the vortex core. 
By interpreting $R$ as the typical local radius of curvature $\ell$ of the
filaments ($\ell \approx L^{-1/2}$, where $L$ is the vortex line density), 
we apply the result to a recent
experiment in which quantum turbulence was visualized.

\section{Equations of motion}

Consider a quantized vortex ring of radius $R$ which moves in the $z$ 
direction in superfluid helium at temperature $T$.  $N$ small buoyant
tracer particles of mass $m$ and radius $a$ are trapped in the vortex
core. For the sake of simplicity we assume that the vortex ring remains
axisymmetric during the evolution and that the particles remain
trapped in it. Let
$(r,\theta,z)$ be cylindrical coordinates.
The vortex position and velocity are
$\rL=(R,0,z)$ and $\vL=(\Rdot,0,\zdot)=(u,0,v)$, where 
$R=R(t)$, $z=z(t)$, $t$ is time, and a dot denotes a
time derivative. Let the circulation vector be
$\bkappa=(0,\kappa,0)$ where $\kappa=9.97 \times 10^{-4}~\rm cm^2/s$,
 and let
$\ukappa=\bkappa/\kappa$ be the unit vector tangent to the ring in
the $\theta$ direction. 
The forces acting on the vortex ring are the Magnus force ${\bf F}_M$, 
the friction force ${\bf F}_D$ with the normal fluid, and the Stokes
force ${\bf F}_S$ induced by the tracer particles. The Magnus
force is ${\bf F}_M=2 \pi R {\bf f}_M$, where
\begin{equation}
{\bf f}_M=\rho_s \bkappa \times (\vL-\vst),
\label{eq:magnus}
\end{equation}
$\rho_s$ is the superfluid density, $\vst=\vs+\vi$ is the total superfluid
velocity at the vortex line, $\vi=(0,0,v_i)$ is the self--induced velocity
of the ring at $T=0$, where
\begin{equation}
v_i=\frac{\kappa}{4 \pi R} \biggl[ \ln{(8R/\xi)}-\frac{1}{2}\biggr],
\label{eq:vi}
\end{equation}
$\xi \approx 10^{-8}~\rm cm$ is the vortex core radius,
and $\vs=(0,0,v_s)$ is an externally applied superflow, which for 
the sake of simplicity we assume in the $z$ direction.
The friction force ${\bf F}_D$ arises from the interaction of phonons
and rotons (which make up the normal fluid)
with the vortex core; we write ${\bf F}_D=2 \pi R {\bf f}_D$, where
the friction per unit length is
\begin{equation}
{\bf f}_D=\rho_s \kappa [\Gamma_0 (\vn-\vL) 
+\Gamma_0' \ukappa \times (\vn-\vL)],
\label{eq:drag}
\end{equation}
and $\vn=(0,0,v_n)$ is an externally applied normal fluid velocity, which
again we assume to be in the $z$ direction. The quantities
$\Gamma_0$ and $\Gamma_0'$ are dimensionless temperature--dependent friction
coefficients related to the friction coefficients
$\gamma_0$ and $\gamma_0'$ calculated by Barenghi,
Donnelly \& Vinen \cite{BDV1983} by
\begin{equation}
\Gamma_0=\frac{\gamma_0}{\rho_s \kappa},
\qquad \qquad
\Gamma_0'=\frac{\gamma_0'}{\rho_s \kappa}.
\label{eq:gammas}
\end{equation}

Note that $\Gamma_0$ and $\Gamma_0'$ can be expressed in terms 
of the more used friction coefficients $B$ and $B'$ as \cite{BDV1983}
\begin{equation}
\gamma_0=\frac{\rho_n \rho}{2 \rho}\kappa 
\frac{B}{[1-B'\rho_n/(2 \rho)]^2 + B^2 \rho_n^2/(4 \rho^2)}, 
\end{equation}
\begin{equation}
\gamma_0'=\frac{\rho_n \rho}{2 \rho}\kappa 
\frac{[B^2 \rho_n/(2 \rho) -B'(1-B'\rho_n/(2 \rho)]}
{[1-B'\rho_n/(2 \rho)]^2 + B^2 \rho_n^2/(4 \rho^2)}.
\end{equation}

If $N$ spherical particles of mass $m=4 \pi a^3 \rho/3$ each, 
where $\rho_p$ is the particle density,
are attached to the vortex ring,
the ring has total mass $mN$ and experiences the Stokes force
\begin{equation}
{\bf F}_S=6\pi a \nu_n \rho_n N (\vn - \vL),
\label{eq:stokes}
\end{equation}
where $\nu_n=\mu/\rho_n$ 
is the normal fluid kinematic viscosity, $\mu$ is the viscosity,
and $\rho_n$ is the normal fluid density. This simple linear Stokes
drag is appropriate because the Reynolds number
based on the particle size is extremely
small. For the sake of simplicity and to make
contact with some experiments~\cite{Paoletti-stats,Bewley-reconnections}
we assume that the particles are buoyant, i.e. they
have density $\rho_p=\rho=\rho_n+\rho_s$, hence there is no Archimedes force.

The equations of motion of the vortex ring are thus
\begin{eqnarray}
\frac{d\rL}{dt}=\vL,\\
mN\frac{d\vL}{dt}=2 \pi R({\bf f}_M+{\bf f}_D)+ {\bf F}_S.
\label{eq:motion1}
\end{eqnarray}

The resulting equations for $R(t)$, $z(t)$, $u(t)=\Rdot$ and $v(t)=\zdot$ are:
\begin{eqnarray}
\Rdot=u,\\
\zdot=v,\\  
\udot=cR [ v-v_s-v_i-\Gamma_0 u +\Gamma_0'(v_n-v) ]
 -\frac{1}{\tau}u,\\ 
\vdot=cR [ - (1- \Gamma_0') u +\Gamma_0 (v_n-v)]
+\frac{1}{\tau}(v_n-v),
\label{eq:motion2}
\end{eqnarray}
where
\begin{equation}
c=\frac{2 \pi \rho_s \kappa}{mN},
\qquad \qquad
\tau=\frac{m}{6 \pi a \nu_n \rho_n}.
\qquad \qquad
\end{equation}
The quantity $\tau$ is the Stokes relaxation time of the particles.

It is convenient to rewrite the equations of motion in dimensionless
form. We choose the initial radius $R_0=R(0)$ as the unit of length, and
$V_0=\kappa/R_0$ as the unit of speed; the unit of time is thus
$R_0^2/\kappa$. We obtain
\begin{equation}
\Rdot=u,
\label{eq:1}
\end{equation}
\begin{equation}
\zdot=v,
\label{eq:2}
\end{equation}
\begin{equation}
v-v_s-v_i-\Gamma_0 u +\Gamma_0'(v_n-v)=
\frac{1}{R}(\epsilon \udot + \zeta u), 
\label{eq:3}
\end{equation}
\begin{equation}
- (1- \Gamma_0') u +\Gamma_0 (v_n-v)=
\frac{1}{R}(\epsilon \vdot + \zeta v),
\label{eq:4}
\end{equation}
where $t$, $R$, $z$, $u$, $v$, $v_i$, $v_n$ and $v_s$ are now
dimensionless. The dimensionless
quantities $\epsilon$ and $\zeta$ are defined as
\begin{equation}
\epsilon=\frac{2N}{3}\left(\frac{a}{R_0}\right)^3
\left(\frac{\rho}{\rho_s}\right),
\qquad \qquad
\zeta=3N\left(\frac{R_0}{a}\right)\left(\frac{\rho_n}{\rho_s}\right) 
\left(\frac{\nu_n}{\kappa} \right).
\end{equation}

The above dimensionless form of the equations of motion allows us
to compare the motion of a vortex ring with $N$ tracers against the motion
of a bare vortex ring ($N=0$). If $N=0$ then $\epsilon=\zeta=0$, and
Eqs.~(\ref{eq:3})-(\ref{eq:4}) reduce to 
\begin{equation}
v-v_s-v_i-\Gamma_0 u +\Gamma_0'(v_n-v)=0,
\label{eq:bare-3}
\end{equation}
\begin{equation}
- (1- \Gamma_0') u +\Gamma_0 (v_n-v)=0.
\label{eq:bare-4}
\end{equation}
The solution is
\begin{equation}
\Rdot=u=\frac{\Gamma_0 (v_n-v_s-v_i)}
{[(1-\Gamma_0')^2+\Gamma_0^2]},
\label{eq:bare-u}
\end{equation}
\begin{equation}
\zdot=v=\frac{(1-\Gamma_0')(v_s+v_i)+v_n[(1-\Gamma_0')\Gamma_0'-\Gamma_0^2]}
{[(1-\Gamma_0')^2+\Gamma_0^2]},
\label{eq:bare-v}
\end{equation}
as found in Ref.~\cite{BDV1983}. If we further set $v_n=v_s=0$ we obtain
\begin{eqnarray}
\Rdot=u=\frac{-\Gamma_0 v_i}{(1-\Gamma_0')^2+\Gamma_0^2},\\ \nonumber
\zdot=v=\frac{(1-\Gamma_0')v_i}{(1-\Gamma_0')^2+\Gamma_0^2},
\label{eq:bare-unforced}
\end{eqnarray}
which means that, as expected, the ring shrinks ($u<0$) and speeds up,
transferring its kinetic energy to the stationary background
normal fluid. If we further set $T=0$ we obtain
the expected stationary solution $u=0$, $v=v_i$.

\section{Results}

We choose values of parameters which are relevant to typical experiments
($T=2~{\rm K}$, $a=10^{-4}~{\rm cm}$ and $R_0=10^{-2}~{\rm cm}$) 
and solve Eqs.~(\ref{eq:1})-(\ref{eq:4})
numerically using the Adams-Bashforth method. 
In all calculations we take $v_n=v_s=0$ for simplicity.
The initial conditions are given by Eqs.~(\ref{eq:bare-u}) and 
(\ref{eq:bare-v}).
During the evolution the vortex ring shrinks. Clearly, when a significant
fraction of the circumference of the ring is covered by
tracers, our model breaks down, as there is not enough vortex length
to provide the correct induced velocity $v_i$. Since we are not
interested in this last stage of the evolution, we arbitrarily
stop our calculations when the tracer particles cover one third of
the circumference,
that is, when the dimensionless ring's radius becomes smaller than
$(3 N /\pi)(a/R_0)$.

\begin{figure}[t]
\includegraphics[width=0.60\linewidth,angle=-90]{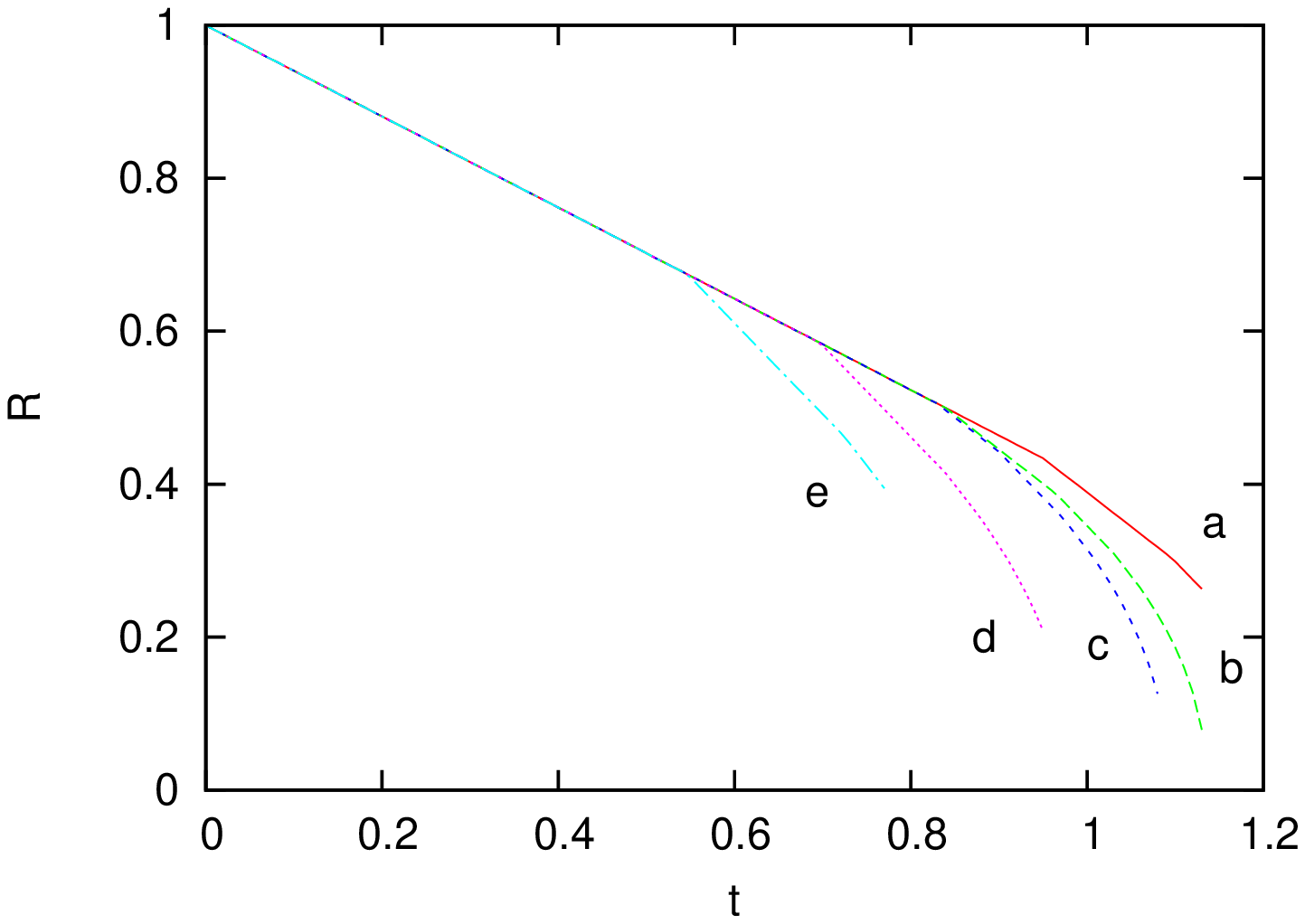}
\caption{(Color online)
Dimensionless radius $R$ of the vortex ring vs time $t$ for:
``a'', solid (red) line: bare vortex ring ($N=0$);
``b'', long-dashed (green) line: vortex ring with $N=10$ tracers, 
$\epsilon=0.19 \times 10^{-5}$, $\zeta=0.34 \times 10^{-1}$;
``c'', short-dashed (blue) line: $N=20$,
$\epsilon=0.37 \times 10^{-5}$, $\zeta=0.68 \times 10^{-1}$;
``d'', dotted (magenta) line: $N=40$,
$\epsilon=0.75 \times 10^{-5}$, $\zeta=0.14$;
``e'', dashed-dotted (cyan) line: $N=80$,
$\epsilon=0.15 \times 10^{-4}$, $\zeta=0.27$.
}
\label{fig1}
\end{figure}

Fig.~\ref{fig1} compares the evolution of the dimensionless radius
$R(t)$ of vortex rings
with $N=10$, $20$, $40$ and $80$ trapped tracer particles to the decay
of a bare vortex ring ($N=0$, solid curve labeled ``a''), 
computed solving Eqs.~(\ref{eq:bare-u}) and (\ref{eq:bare-v}). 
It is apparent that the more particles are trapped (hence the
larger $\epsilon$ and $\zeta$ are), the faster the decay of
the ring is compared to the decay of a bare ring.

\begin{figure}[t]
\includegraphics[width=0.60\linewidth,angle=-90]{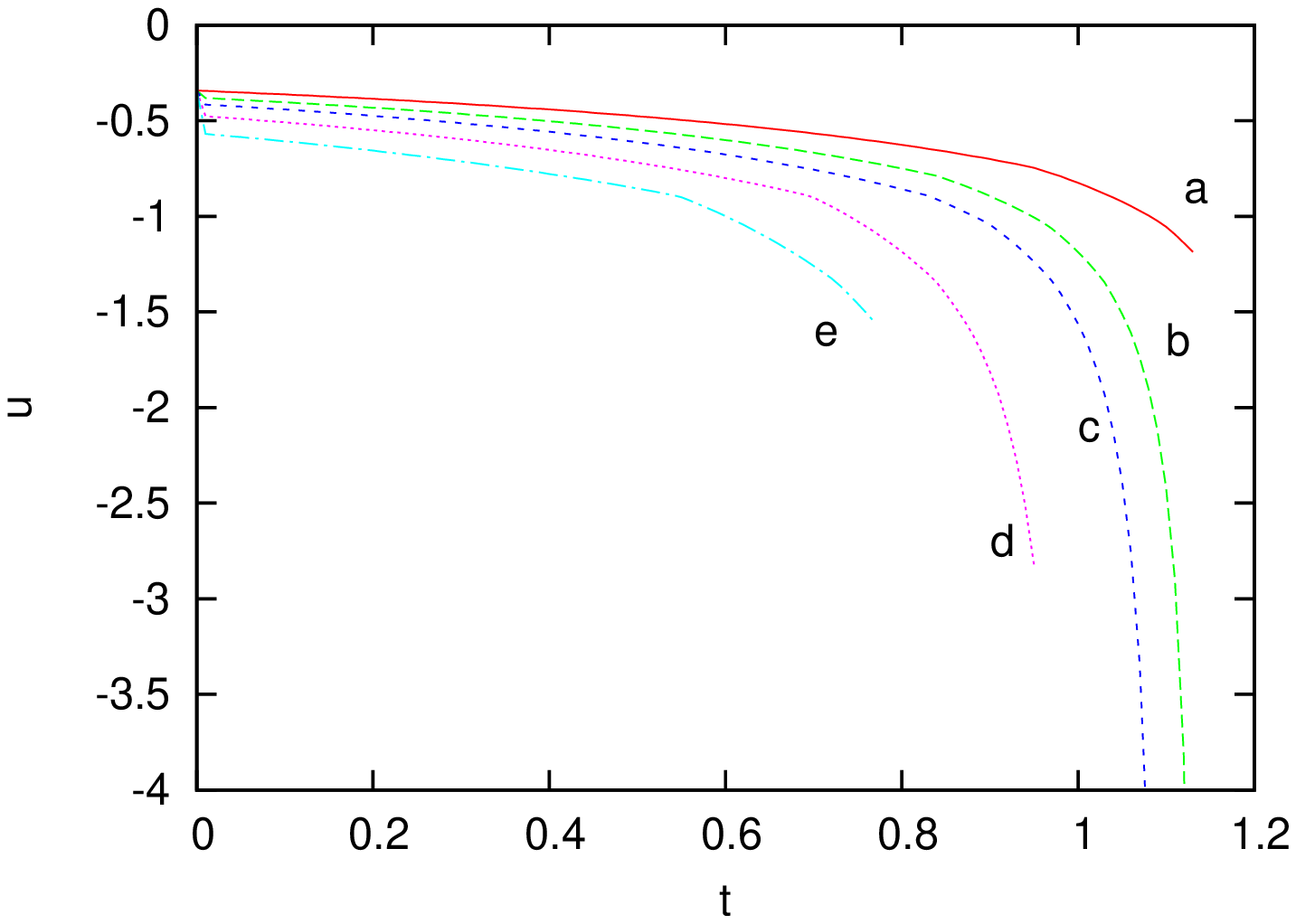}
\caption{(Color online)
Dimensionless radial velocity $u$ of vortex ring vs time $t$
for $N=0$ (bare ring, solid line labeled ``a'') and $N=10$, $20$, $40$ and 
$80$ tracers. (Labels and line styles (colors) correspond to those of
Fig.~\ref{fig1}.)
}
\label{fig2}
\end{figure}

Fig.~\ref{fig2} plots the corresponding time dependence of the radial velocity
component $u=\Rdot$. Initially the ring shrinks at
constant rate, then the radius decreases faster and faster. 

\begin{figure}[t]
\includegraphics[width=0.60\linewidth,angle=-90]{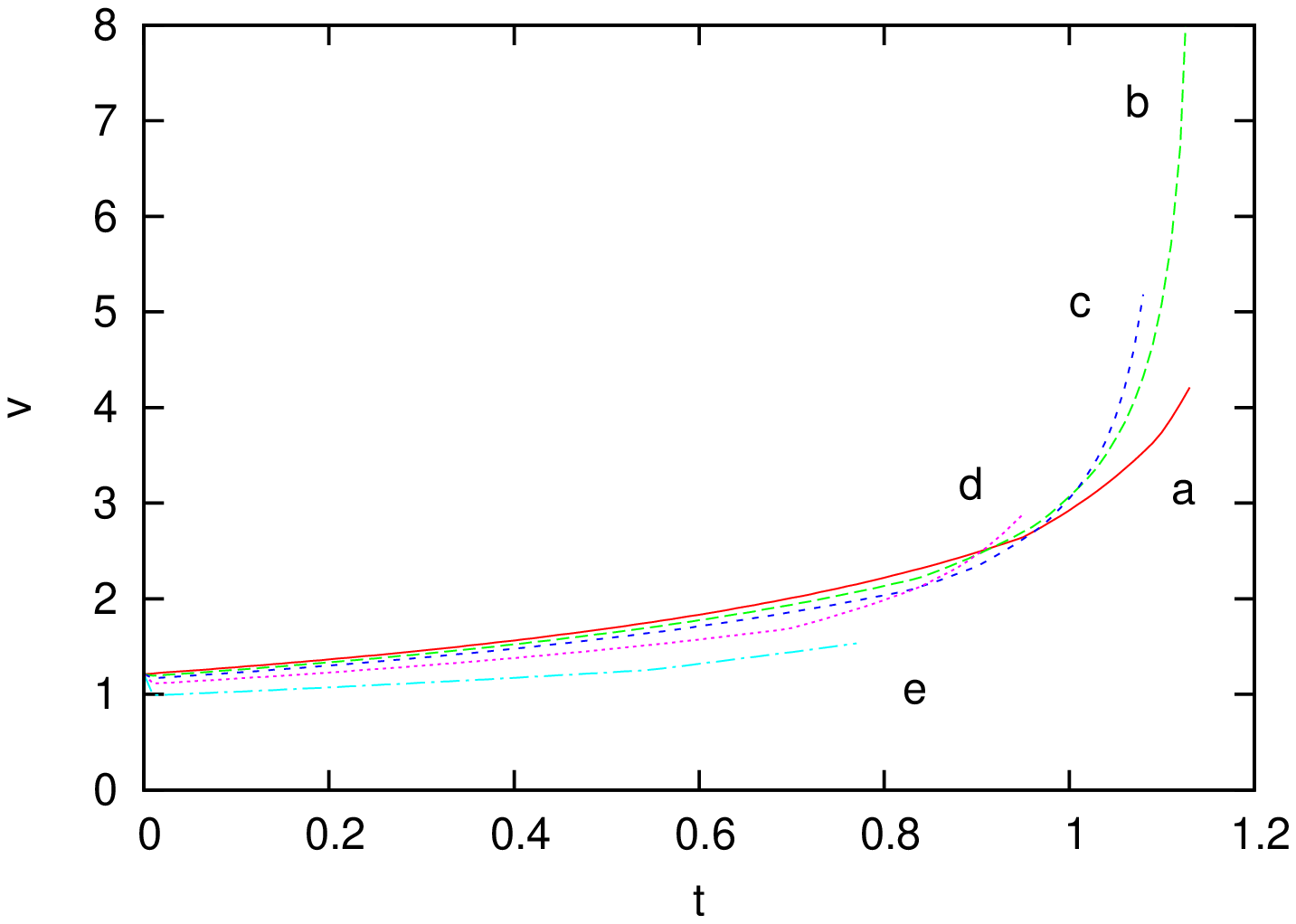}
\caption{(Color online)
Dimensionless axial velocity $v$ of vortex ring vs time $t$ 
for $N=0$ (bare ring, solid (red) line labeled ``a'') 
and $N=10$, $20$, $40$ and 
$80$ tracers.
Labels and line styles (colors) correspond to those of
Figs.~\ref{fig1} and \ref{fig2}.
}
\label{fig3}
\end{figure}

The axial velocity component $\zdot=v$ is shown in Fig.~\ref{fig3}.
Initially $v$ remains similar to the
velocity of the bare ring; during the
final part of the evolution a vortex ring with tracers moves
much faster than a bare vortex ring, because, as shown in Fig.~\ref{fig1},
the radius has become much smaller.

\begin{figure}[t]
\includegraphics[width=0.60\linewidth,angle=-90]{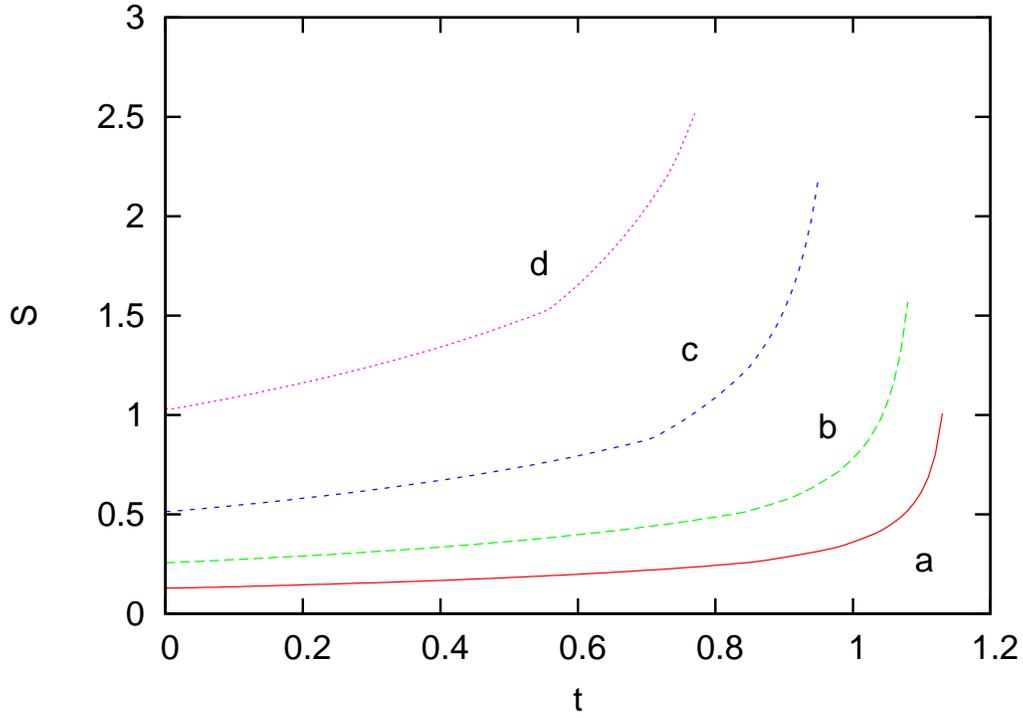}
\caption{(Color online)
Superfluid Stokes number vs dimensionless time $t$ 
for $N=10$, $20$, $40$ and $80$ tracers. 
Labels and line styles (colors) correspond to those of
Figs.~\ref{fig1}-\ref{fig3}.
}
\label{fig4}
\end{figure}

The results described above suggest that the evolution of
vortex rings which contain tracers is sufficiently similar to the
evolution of a bare vortex ring, provided that the ring is not too small
and that it does not contain too many tracers. The important question
is how to quantify the perturbation which the tracers induce on the
the ring's evolution so that one can decide whether the visualization with
tracers affects or not the evolution of vortices. 
To answer this question we start by remarking 
that the ring's decay is caused by both friction
and Stokes forces; going back to dimensional variables, we note
that the ratio of the magnitudes of these forces is 
\begin{equation}
S=\frac{F_S}{F_D}=\frac{3N}{\Gamma_0}\left(\frac{a}{R}\right)
\left(\frac{\nu_n}{\kappa}\right) \left(\frac{\rho_n}{\rho_s}\right)
=\frac{\zeta}{\Gamma_0}\left(\frac{R_0}{R}\right).
\label{eq:s}
\end{equation}
We call $S$ the superfluid Stokes number.
In the absence of external forcing ($v_n=v_s=0$)
the superfluid Stokes number decreases during the evolution,
starting from the initial value $S(0)=\zeta/\Gamma_0$ at $t=0$,
because $R$ decreases with time. Fig.~\ref{fig4} shows the time evolution
of $S$ corresponding to Figs.~\ref{fig1}, \ref{fig2}, and \ref{fig3}.
It is apparent that if $S\ll1$ the evolution of a vortex ring with
tracers is very similar to that of a bare vortex ring, but if $S$ becomes
of order unity or larger then the tracers affect the ring's motion
significantly.

\begin{figure}[t]
\includegraphics[width=0.60\linewidth,angle=-90]{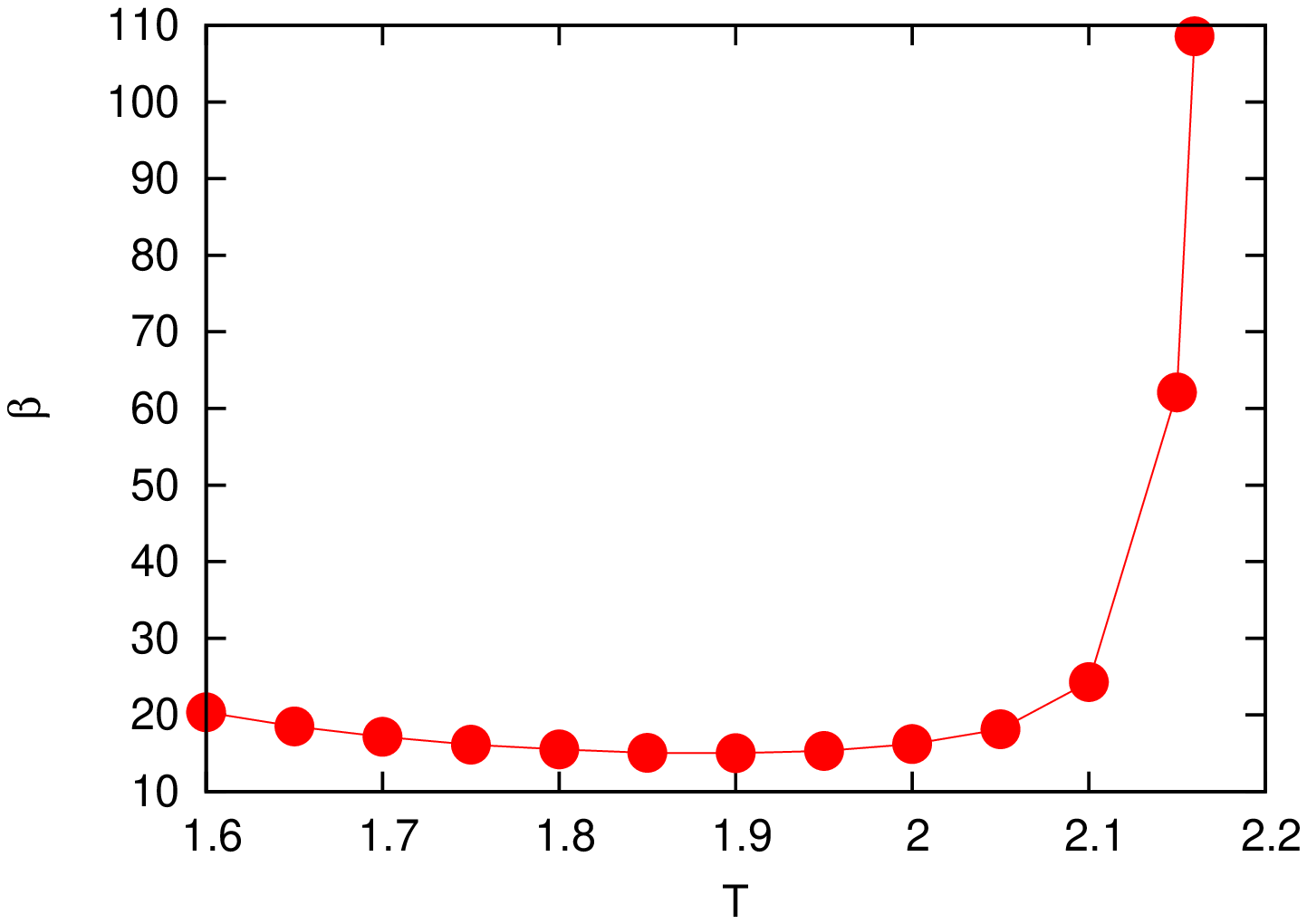}
\caption{(Color online)
Dimensionless temperature dependent prefactor $\beta$ vs temperature $T$
($\rm K$).
}
\label{fig5}
\end{figure}

To put in evidence the temperature dependence of the superfluid Stokes
number, we rewrite Eq.~(\ref{eq:s}) as
\begin{equation}
S=\delta \beta,
\label{eq:s2}
\end{equation}
where 
\begin{equation}
\delta=\frac{Na}{2\pi R}
\label{eq:delta}
\end{equation}
is the distance between tracers along 
the circumference of the ring in units of tracer size, and
\begin{equation}
\beta=\frac{6 \pi}{\Gamma_0}\left(\frac{\nu_n}{\kappa}\right)
\left(\frac{\rho_n}{\rho_s}\right)
\label{eq:beta}
\end{equation}
is a strongly temperature dependent prefactor.
Fig.~\ref{fig5} shows that $\beta$ is approximately constant
for temperatures below $2~{\rm K}$ but it increases sharply
above  $2\rm~K$. This means that, for
the same geometry (ring's radius and number of tracers along the ring),
the superfluid Stokes number is bigger for $T>2\rm~K$, 
hence the motion of vortex rings is
much more disturbed by the presence of tracer particles than shown
for example in Figs.~\ref{fig1}-\ref{fig3}.

\section{Application to quantum turbulence}

Isotropic homogeneous quantum turbulence
is characterized by the vortex
line density $L=\Lambda/V$, where $\Lambda$ is the total
vortex length and $V$ is the volume. In the first approximation
this form of turbulence can be generated by applying a heat flux
\cite{Tough}.  From the value of $L$ one infers
that the average vortex separation and the typical radius of curvature
are of the order of $\ell \sim L^{-1/2}$.
In a recent experiment~\cite{Paoletti-JPSJ}, solid hydrogen particles were
used to visualize such turbulence: some tracers moved along the normal fluid
in the direction of the heat flux, and other tracers were trapped on the 
quantized vortices as the vortex tangle drifted in the opposite direction. 
At $T=2\rm~K$ the turbulence  generated by the heat flux 
$\dot Q =90~{\rm mW/cm^2}$ corresponds to
approximately~\cite{note2}
$\ell \approx 0.006\rm~cm$. The hydrogen volume fraction
used in the experiment was $\phi=V_{H_2}/V \approx 10^{-7}$.
Generalizing from a vortex ring of length $2 \pi R$ to a vortex tangle
of length $\Lambda=LV$ and using $V_{H_2}=4N \pi a^3/3$,
the superfluid Stokes number can be estimated as
\begin{equation}
S=\delta \beta \approx 
\frac{3 \phi \beta}{4 a\pi}\left(\frac{\ell}{a}\right)^2,
\end{equation}
where $\delta=Na/(2 \pi R) \approx (N/V)(V/\Lambda)a=(N/V)(a/L)$.
If one tenth of the tracers are trapped inside vortices, using
$\beta \approx 20$ (see Fig.~\ref{fig5}) we obtain
$S \approx 5 \times 10^{-4}\ll1$. We conclude that in this particular
experiment the tracers
have not disturbed the dynamics of the vortex tangle. 

\section{Discussion}

We have derived the equations of motion of a vortex ring which contains
$N$ buoyant tracer particles trapped on the vortex core. We have shown
typical solutions of these equations to illustrate the difference
between the motion of a bare, undisturbed vortex ring, and that of
a vortex ring which is visualized by tracer particles.

A first approximate measure of the disturbance caused by trapped
particles is the geometrical quantity $\delta$ given by Eq.~(\ref{eq:delta}),
the separation between
tracers along the ring in unit of tracer's size. Clearly $\delta$ must
be small for tracers not to disturb the vortex. 
A more precise measure is the superfluid Stokes number $S=\delta \beta$,
where the dimensionless temperature dependent quantity $\beta$ is
approximately constant for $T<2\rm~K$ but becomes very
large for $T>2\rm~K$. 

Finally, we have shown how this criterion is relevant to 
quantum turbulence, and can be used to determine
the maximum hydrogen volume factor which can be used to visualize 
the turbulent flow without disturbing it.
We have applied the criterion to a recent 
experiment~\cite{Paoletti-JPSJ} in which the drift
of vortex lines generated by a heat flux was measured and found
that the disturbance was negligible. It would be interesting
to apply the same criterion to experiments in which vortex reconnections
were detected \cite{Bewley-reconnections}: this case is more
challenging, as during the reconnection process much smaller
values of the radius of curvature are expected to arise which may
be disturbed by tracer particles.

\section{Acknowledgements}

We are grateful to W.~F.~Vinen. S.~W.~Van Sciver, M.~S.~Paoletti, 
D.~P.~Lathrop,
and K.~R.~Sreenivasan for fruitful discussions.

\vfill
\eject

\newpage


\newpage

\vfill
\eject

\vfill
\eject

\vfill
\eject

\vfill
\eject

\begin{thebibliography}{99}
\bibitem{note2}
The heat flux $\dot Q$ induces the counterflow velocity 
$v_{ns}=v_n-v_s=\dot Q/(\rho_s S' T)$ where $S'$ is the entropy.
At $T=2\rm~K$ $S'=0.956~\rm J/(g~K)$ and $\rho_s=0.065~\rm g/cm^3$,
so $v_{ns}=0.73~\rm cm/s$. The vortex line density is thus
$L=\gamma^2 v_{ns}^2 \approx 2.7 \times 10^4~\rm cm^{-2}$,
where, following Ref.~\cite{Tough}, $\gamma\approx 220~\rm s/cm^2$.
\end{thebibliography}
\end{document}